\documentstyle[epsfig]{article}
\def\Journal#1#2#3#4{{#1} {\bf #2}, #3 (#4)}

\def\NPA{{\em Nucl. Phys.} A}

\def\PLB{{\em Phys. Lett.} B}

\def\PREP{\em Phys. Rep.}

\def\PRC{{\em Phys. Rev.} C}

\def\ANNP{\em Ann. Phys. (N.Y.)}
\def\r{\vec r}

\def\p{\vec p}

\def\q{\vec q}
\def\ss{\mbox{\boldmath $\sigma$}}

\begin{document}

\title{
EXPLORING NUCLEON-NUCLEON CORRELATIONS IN $(e,e'NN)$ REACTIONS
}

\author{H.\ M\"UTHER\\
Institut f\"ur Theoretische Physik,\\Universit\"at T\"ubingen, Germany}

\maketitle

\begin{abstract} 
Correlations in the nuclear wave-function beyond the mean-field or 
Hartree-Fock approximation are very important to describe basic properties of
nuclear structure. Attempts are made to explore details of these correlations in
exclusive nucleon knock-out by electron scattering experiments. Basic results of
$(e,e'p)$ experiments are reviewed. The role of correlations in $(e,e'NN)$
experiments is discussed. Special attention is paid to a consistent description
of the competing effects due to final state interaction, meson exchange current
and isobar currents. Results are discussed for systematic studies of these
features in nuclear matter as well as for specific examples for the finite
nucleus $^{16}$O. 
\end{abstract}

\section{NN Correlations}

Nuclei are a very intriguing object to explore the theory of quantum-many-body
systems. One of the reasons is that realistic wave functions of nuclear systems
must exhibit strong two-particle correlations. This can be demonstrated in a
little `theoretical experiment': Assuming a realistic model for the
nucleon-nucleon (NN) interaction\cite{cdb,argv18,nijm1,bonc,reid}, 
this means an interaction which reproduces the empirical data of NN scattering 
below the pion threshold, one may calculate the energy of nuclear matter 
within the mean field or Hartree-Fock approximation.
Results of such a calculation are listed in the first row of table \ref{tab1}. 
One finds that all these interactions yield a positive value for the energy per
nucleon, which means that nuclear matter as well as all nuclei would be unbound.
Only after the effects of two-body correlations are included, one obtains a 
value 
which is in rough agreement with the empirical value of -16 MeV per nucleon.
This demonstrates that nuclear correlations are indispensable to describe the
structure of nuclei. 

\begin{table}[t]
\begin{center}
\begin{tabular}{|c|rrr|rr|}
\hline
& CDB & ArgV18 & Nijm1 & Bonn C & Reid \\
\hline
$E_{HF}$ & 4.64 & 30.34 & 12.08 & 29.56 & 176.20 \\
$E_{Corr}$ & -17.11 & -15.85 & -15.82 & -14.40 & -12.47 \\
$V_{\pi HF}$ & 16.7 & 15.8 & 15.0 & 17.8 &\\
$V_{\pi Corr}$ & --2.30 & -40.35 & -28.98 & -45.74 & \\
$T$ & 36.23 & 47.07 & 39.26 & 40.55 & 49.04 \\
\hline
\end{tabular}
\end{center}
\caption{Energy per nucleon for nuclear matter at the empricial saturation
density. Results are displayed for the NN interactions CDB \protect\cite{cdb},
ArgV18 \protect\cite{argv18}, Nijm1 \protect\cite{nijm1}, Bonn C
\protect\cite{bonc} and the Reid potential \protect\cite{reid}. The results
obtained in the Hartree-Fock approximation $E_{HF}$ are compared to those of
Brueckner-Hartee-Fock calculations ($E_{Corr}$). Furthermore the contribution of
the $\pi$ exchange to the total energy in Born approximation ($V_{\pi HF}$) and
including the effects of correlations ($V_{\pi Corr}$) as well as the
expectation value of the kinetic energy ($T$) are listed. All entries are given
in MeV.}
\label{tab1}
\end{table}

In order to explore dominant components of these correlation, table \ref{tab1}
also lists the expectation value of the $\pi$-exchange contribution to the NN
interaction using the HF approximation ($V_{\pi HF}$) and with inclusion of the    
correlation effects ($V_{\pi Corr}$). One finds that the gain in binding energy
is not only due to the central short-range correlation effects, i.e.~the nuclear
wave function tries to minimize the probability that two nucleons approach each
other so close that they feel the repulsive core of the interaction. A large
part of this gain in binding energy is due to pionic correlations which are
dominated by the effects of the tensor force. 

The different interaction models all reproduce the same empirical NN scattering
phase shifts. This is true in particular for the modern NN interactions: the
charge-dependent Bonn potential (CDB)\cite{cdb}, the Argonne V18
(ArgV18)\cite{argv18} and the Nijmegen interaction (Nijm1)\cite{nijm1}, which
all yield an excellent fit of the same phase shifts. Nevertheless, they
predict quite different correlations. This can be seen e.g.~from inspecting the
expectation values for the kinetic energies per nucleon (denoted as $T$ in table
\ref{tab1}). This means that correlations are a significant fingerprint of the
interaction of two nucleons in a nuclear medium. So if we find a way to measure
details of these correlations, we shall obtain information on the validity of
the various models for the NN interaction.

\section{Correlations and exclusive $(e,e'p)$ reactions}

The uncorrelated Hartree-Fock state of nuclear matter is given as a Slater
determinant of plane waves, in which all states with momenta $k$ smaller than
the Fermi momentum $k_F$ are occupied, while all others are completely
unoccupied. Correlation in the wave function beyond the mean field approach will
lead to occupation of states with $k$ larger than $k_F$. Therefore correlations
should be reflected in an 
enhancement of the momentum distribution at high momenta. Indeed, microscopic
calculations exhibit such an enhancement for nuclear matter as well as for 
finite nuclei\cite{p1,p2}. One could try to measure this momentum distribution
by means of exclusive $(e,e'p)$ reactions at low missing energies, such that
residual nucleus remains in the ground state or other well defined bound state.
From the momentum transfer $q$ of the scattered electron and the momentum $p$ of
the outgoing nucleon one can calculate the momentum of the nucleus before the
absorption of the photon and therefore obtain direct information on the momentum
distribution of the nucleons inside the nucleus. 

This idea, however, suffers from a little inaccuracy. To demonstrate this we
write the momentum distribution $n(k)$ representing the ground state wave
function of the target nucleus by $\Psi_A$, denoting the creation (annihilation)
operator for a nucleon with momentum $k$ by $a^\dagger_k$ ($a_k$), as
\begin{eqnarray*}
n(k) & = & < \Psi \vert a^\dagger_k a_k \vert \Psi > \\
& = & \int_0^\infty dE \, < \Psi \vert a^\dagger_k \vert \Phi_{A-1} (E) >< 
\Phi_{A-1} (E)\vert a_k \vert \Psi >\\
& = & \int_0^\infty dE \, S(k,E) \\
\mbox{with} S(k,E) & = & \left| < \Psi \vert a^\dagger_k \vert \Phi_{A-1} (E) >
\right|^2\, .
\end{eqnarray*}
 In the second line of this equation we have inserted the complete set of
eigenstates for the residual nucleus with $A-1$ nucleons and excitation energy
$E$. Therefore, if one performs an exclusive $(e,e'p)$ experiment leading to the 
residual nucleus in its ground state, one does not probe the momentum
distribution but the spectral function at an energy $E=0$. While the total
momentum distribution exhibits the enhancement at high momenta discussed above,
the spectral function at small energies does not have this feature\cite{p1} and
the momentum distribution extracted from such experiments is very similar to the
one derived from Hartee-Fock wave functions.

\begin{figure}[t]
\begin{center}
\begin{minipage}[t]{8 cm}
\epsfig{file=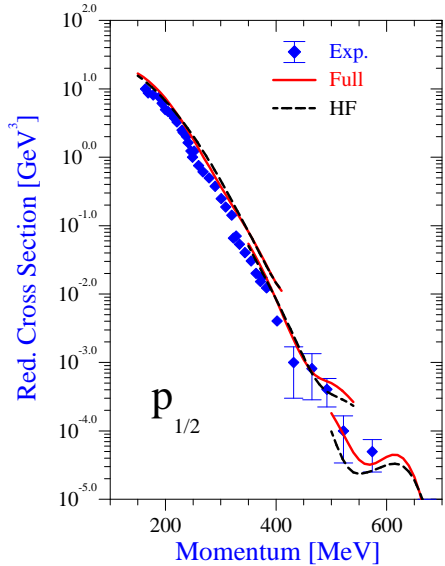,scale=0.5,angle=270}
\end{minipage}
\end{center}
\caption{Reduced cross section for the $^{16}$O($e,e'p$)$^{15}$N
reaction leading to the ground state ($1/2^-$) of $^{15}$N in the 
kinematical conditions considered in the
experiment at MAMI (Mainz) \protect\cite{mainz}. Results for the
mean-field description (HF) and the fully correlated spectral function
(Full) are presented.}
\label{fig1}
\end{figure}

This is demonstrated by Figure \ref{fig1}, which compares experimental data of
$(e,e'p)$ experiments on $^{16}O$ leading to the ground state of the residual
nucleus $^{15}N$, which were performed at MAMI in Mainz\cite{mainz}, with
theoretical calculations\cite{nili}. The calculation account for the final state
interaction of the outgoing nucleon with the residual nucleus by means of a
relativistic optical potential. One finds that the spectral function
calculated with inclusion of correlation yields the same shape as the
corresponding Hartree-Fock approximation. The only difference being the global
normalization: the spectroscopic factor. 

Therefore exclusive $(e,e'p)$ reactions yield a rather limited amount of
information on correlation effects, they are sensitive to the mean field
properties of the nuclear system. There is a major discussion of these mean
field properties in nuclear physics: Motivated by the success of the Walecka
model\cite{walecka}, attempts have been made to include relativistic features in
microscopic nuclear many-body studies. Such attempts are often referred to as
Dirac-Brueckner-Hartree-Fock calculations\cite{bro1,bro2}. The main prediction
of these relativistic nuclear structure calculations is that the small component
of the Dirac spinors for the nucleon inside a nucleus is enhanced relative to
the small component of a free nucleon with the same momentum. This enhancement
can be parameterized in terms of an effective Dirac mass $m^*$ which is
significantly smaller than the bare nucleon mass. 

\begin{figure}[t]
\begin{center}
\begin{minipage}[t]{10 cm}
\epsfig{file=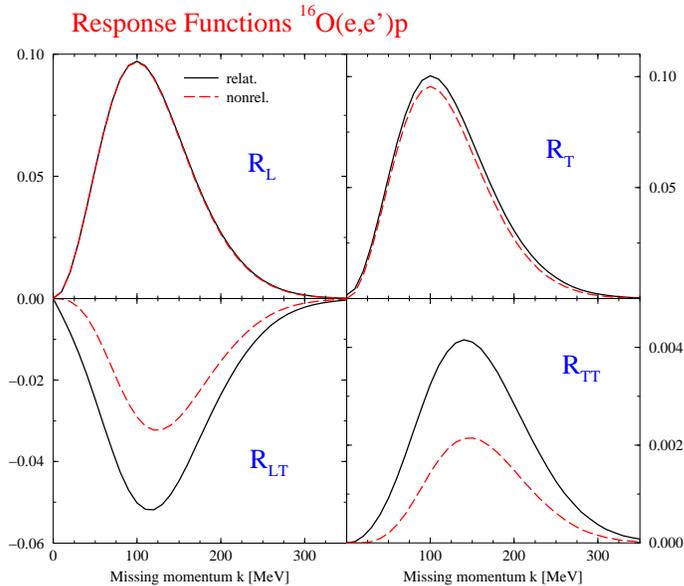,scale=0.4,angle=270}
\end{minipage}
\end{center}
\caption{Response functions $R_L$, $R_T$,
$R_{LT}$ and $R_{TT}$ for the knockout of a nucleon from $p_{1/2}$ state in
$^{16}O$ as a function of the missing momentum. Comparison of relativistic and
non-relativistic approach}
\label{fig2}
\end{figure}

Can one observe this enhancement of the small component of the Dirac spinor by
means of $(e,e'p)$ experiments? Theoretical calculations predict that this may
be possible, if one performs a more detailed analysis of the corresponding cross
section. For that purpose one decomposes the cross section into a contraction of
hadronic responses and the appropriate electron
contributions, which are defined as in \cite{Kell96}
$$
\frac{m\vert {p}_x \vert}{(2\pi)^3}\;\sigma_{Mott}
\left(V_LR_L+V_TR_T
+V_{LT}R_{LT}\cos{\phi}+V_{TT}R_{TT}
\cos{2\phi}\right)\;.
$$
Results for the hadronic response functions with and without the relativistic
effect\cite{ulrich} are displayed in Figure \ref{fig2}. While the relativistic
features do not effect the longitudinal $R_L$ and transverse repons functions
$R_T$, they predict an enhancement of the interference structure functions
$R_{LT}$ and $R_{TT}$ as compared to the non-relativistic reduction. This
feature is discussed more in detail by Moya de Guerra and
Udias\cite{udias}. First experimental results on $R_{LT}$ are presented by
Bertozzi\cite{bill} at this workshop.  

\section{Kinematical study of $(e,e'NN)$ in nuclear matter}

As exclusive one-nucleon knock-out experiments only yield limited information on
NN correlations, one may try to investigate exclusive $(e,e'NN)$ reactions,
i.e.\ triple coincidence experiments in which the energies of the two outgoing
nucleons and the energy of the scattered electron guarantee that the rest of the
target nucleus remains in the ground state or a well defined excited state.
The idea that processes in which the virtual photon, produced by the scattered
electron, is absorbed by a pair of nucleons should be sensitive to the
correlations between these two nucleons.

\begin{figure}[th]
\begin{center}
\epsfig{file=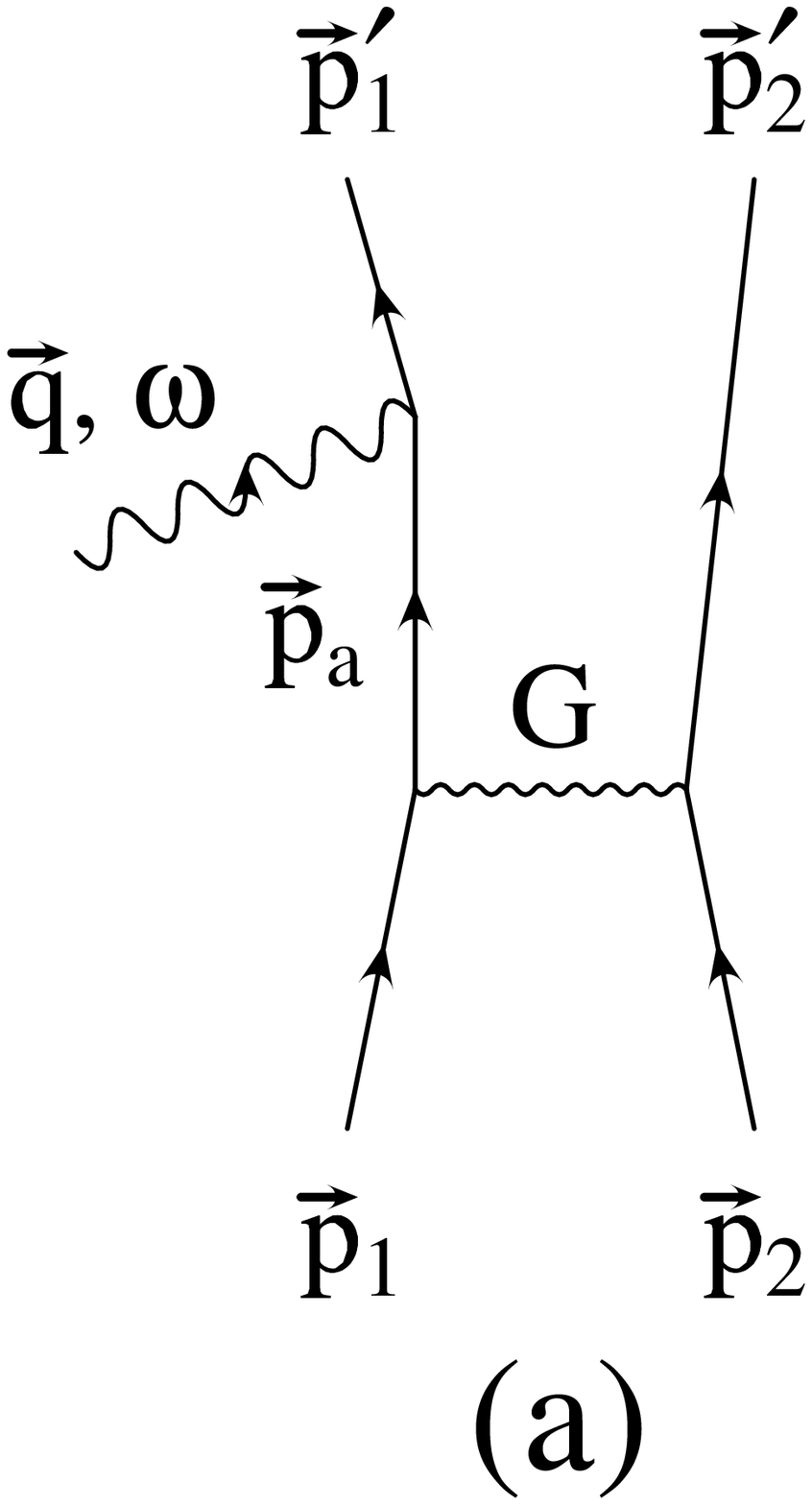,scale=0.18}\hspace{.5cm}
\epsfig{file=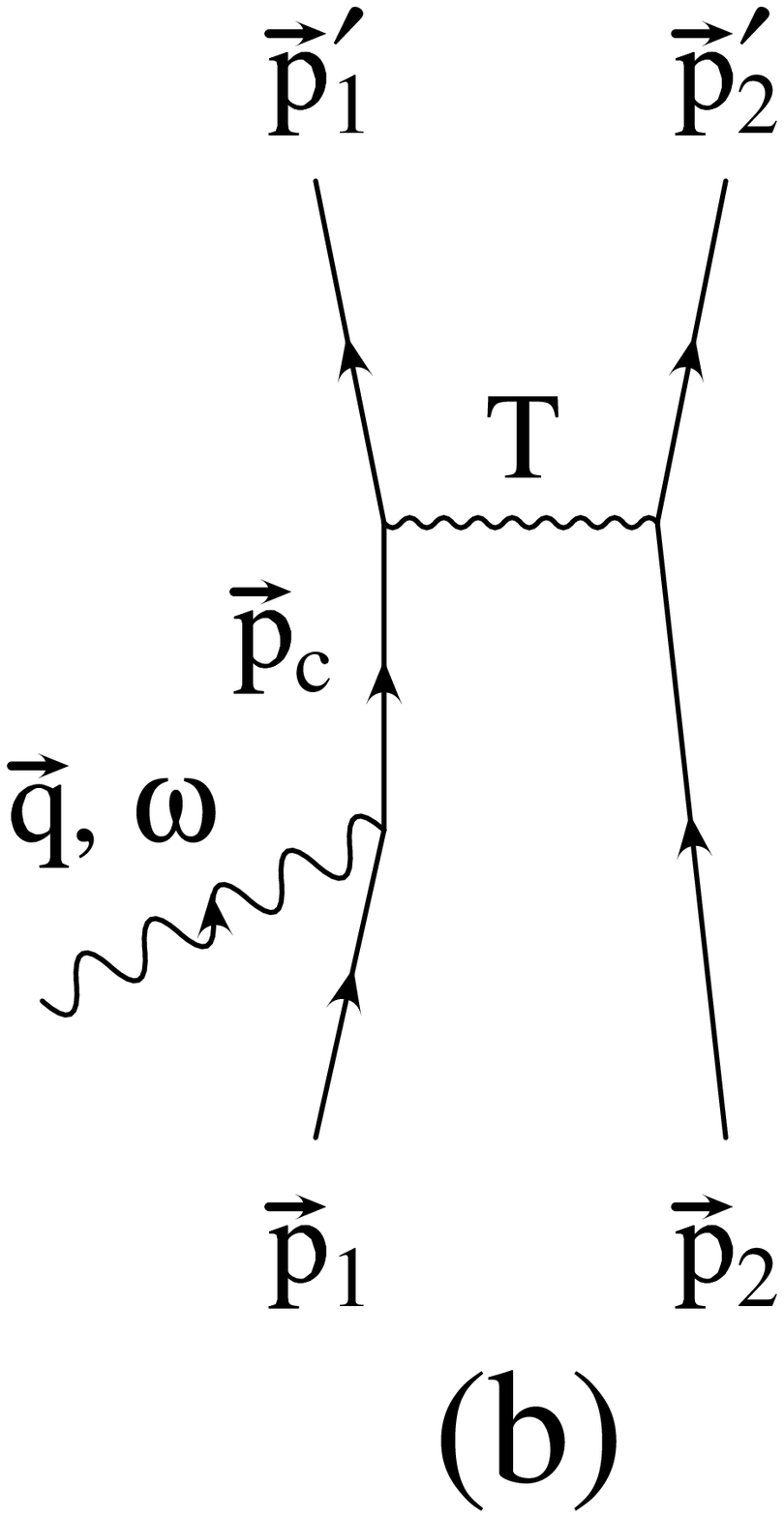,scale=0.18}\hspace{.5cm}
\epsfig{file=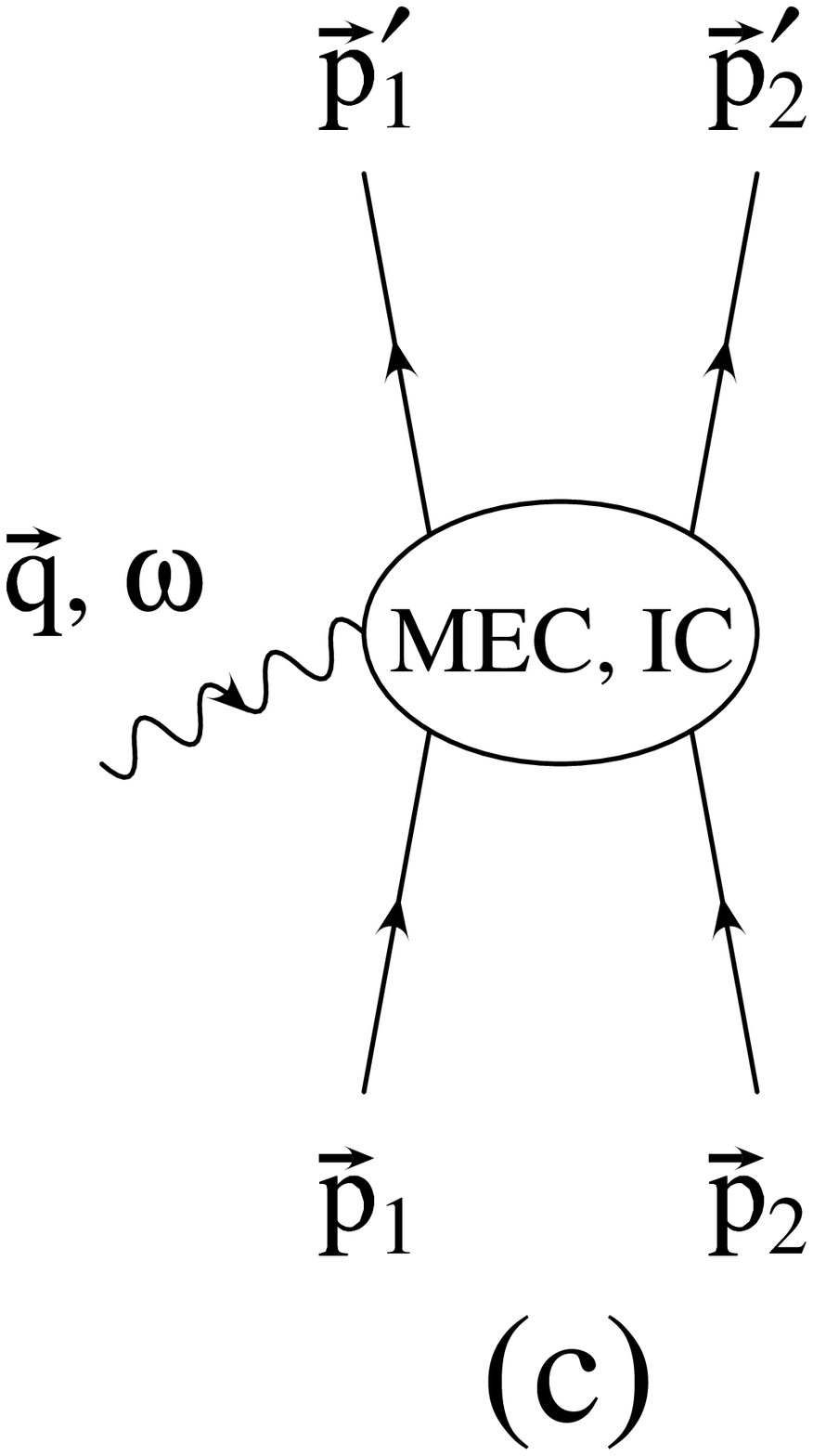,scale=0.18}
\end{center}
\caption{\label{fig3}
Diagrams for the different processes contributing to the $(e,e'2N)$
reaction. Diagram (a) and (b) show the absorption of the photon by a
single nucleon. The nucleon-nucleon correlations are described by the $G$
matrix. Diagram (c) depicts photon absorption via meson exchange (MEC) or
isobaric currents (IC)}
\end{figure}

Unfortunately, however, this process which is represented by the diagram
in Figure \ref{fig3}a, competes with the other processes described by the 
diagrams of Figure \ref{fig3}b and c. These last two diagrams refer to the
effects of final-state-interaction (FSI) and contributions of two-body currents.
Here we denote by final state interaction not just the feature that each of the
outgoing nucleons feels the remaining nucleus in terms of an optical potential.
Here we call FSI the effect, that one of the nucleons absorbs the photon,
propagates (on or off-shell) and then shares the momentum end energy of the
photon by interacting with the second nucleon which is also knocked out the
target. The processes described in Figures \ref{fig3}a and \ref{fig3}b,
correlations and FSI, are rather similar, they differ only by the time ordering
of NN interaction and photon absorption. Therefore it seems evident that one
must consider both effects in an equivalent way. Nevertheless, all studies up to
now have ignored this equivalency but just included the correlation effect in
terms of a correlated two-body wave function. In our approach we will assume the
same interaction to be responsible for the correlations and the FSI,
correlations are evaluated in terms of the Brueckner G-matrix\cite{knoed}, 
while the T-matrix derived from the very same interaction is used to determine 
FSI.

\begin{figure}[bt]
\hspace{1cm}\epsfig{file=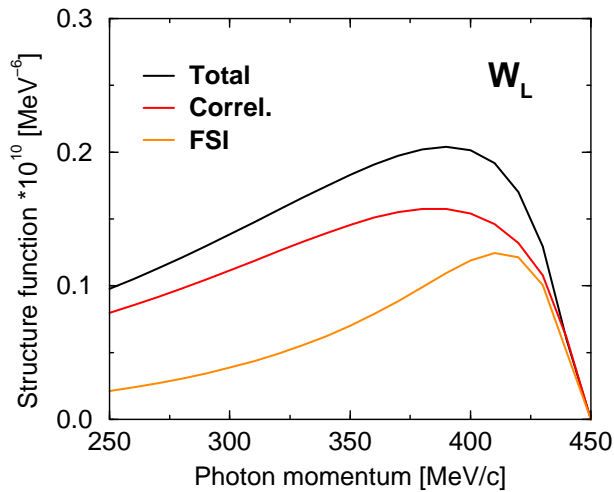,scale=0.45}
\caption{\label{fig4}
Longitudinal structure function
for the knockout of a proton-proton pair in a 'super parallel' kinematical
situation with angles $\theta_{p,1}'=0^{\rm o}$ and
$\theta_{p,2}'=180^{\rm o}$ of the two protons with respect to the
direction of the photon momentum. The figure has been generated assuming
final kinetic energies $T_{p,1}=156\,{\rm MeV}$ and $T_{p,2}=33\,{\rm
MeV}$ of the two protons. The structure function is displayed as a function of
the photon momentum, keeping
the photon energy constant at $\omega=215\,{\rm MeV}$}
\end{figure}

The two-body current contributions of Figure \ref{fig3}c include Meson Exchange
Current (MEC) and Isobar Current (IC) contributions. The MEC effects should be
calculated consistently with the meson exchange terms included in the NN
interaction, used to calculate correlations and FSI. In our calculations up to
day we only account for the contributions die to the exchange of the pions. Note
that the pion-seagull and pion in flight term only contribute is the emitted
pair contains a proton and a neutron. Contributions of other charged mesons like
e.g. the $\rho$ meson have been considered e.g.~by Vanderhaeghen et al.
\cite{janrho} and shall also be included in future investigations.

The IC contributions contain diagrams like the ones displayed in Figures
\ref{fig3}a and b. The only difference being that the intermediate nucleon line
is replaced by the propagation of the $\Delta$ excitation. This demonstrates
that also IC contributions should be treated in terms if baryon-baryon
interactions accounting for admixture of $\Delta$ configurations to the target
wave functions as well as FSI effects with intermediate isobar terms. Presently
the IC terms are evaluated in terms of the Born diagrams, including again only
$\pi$ exchange for the transition interactions $NN \iff N\Delta$. Also in this
case one should account for the effects of the $\rho$ exchange.  

In this section I would like to present results for the various contributions
just introduced, calculated for nuclear matter at saturation density. Of
course, this study will not lead to any result, which can directly be compared
with experimental data for a specific target nucleus. The idea is to get some
general features, which are independent on the specific target nucleus or final
state of the residual nucleus. We would like to see, if we can provide general
information about the importance of the various contributions just discussed. It
is the hope, that one may find special kinematical situations, 
in which one of the
contributions mentioned above is dominating over others. All results discussed
here have been obtained with the Bonn A potential defined by
Machleidt\cite{bonc}. Details of these calculations are described in
reference\cite{knoed}.

As a first example we consider the longitudinal structure function
for the knockout of a proton-proton pair. One of the protons is emitted parallel
to the momentum of the virtual photon with an energy of $T_{p,1}=156\,{\rm
MeV}$, while the second is emitted antiparallel to the photon momentum with an
energy of $T_{p,2}=33\,{\rm MeV}$ (see Figure \ref{fig4}). 
This is called the `super-parallel kinematic',
which should be appropriate for a separation of longitudinal and transverse
structure functions. In this situation the dominant contribution to the
longitudinal response function is due to correlation effects (red curve). But
also the FSI effects contribute in a non-negligible way to the cross section
(yellow curve), although the two protons are emitted in opposite directions. 

\begin{figure}[t]
\hspace{1cm}\epsfig{file=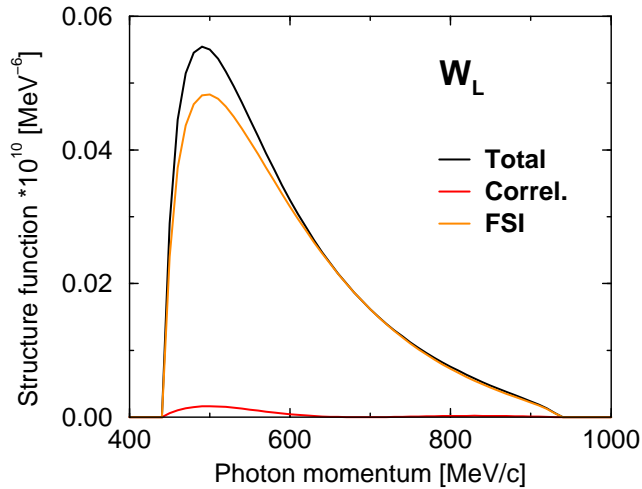,scale=0.45}
\caption{\label{fig5}
Longitudinal structure function for ($e,e'pp$) with $\omega=230\,{\rm MeV}$,
assuming $\theta_{p_i}=\pm 30^{\rm o}$ and $T_{p,i}=70\,{\rm MeV}$}.  
\end{figure}

The effects of FSI are much more important, if we request that the two protons
are emitted in a more symmetric way. As an example we show the longitudinal
structure function for ($e,e'pp$), requesting that each of the protons carries
away an energy of 70 MeV and is emitted with an angle of $30^{\rm o}$ or 
$-30^{\rm o}$ with respect to the momentum transfer $q$ of the virtual photon.
Corresponding results are displayed in Figure \ref{fig5}. For this kinematical
situation the FSI contribution is much more important than the correlation
effect.

\begin{figure}[tb]
\hspace{1cm}\epsfig{file=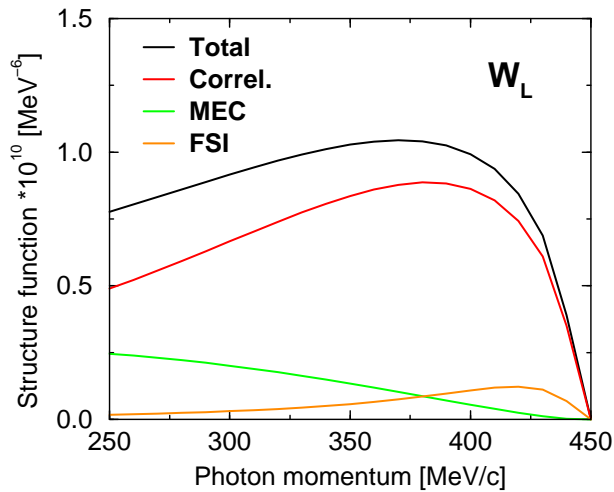,scale=0.45}
\caption{\label{fig6}
Longitudinal structure function for ($e,e'pn$) for the kinematical condition as
displayed in Figure 4}
\end{figure}

As a last example we present the results for the longitudinal structure function
for the ($e,e'pn$) reaction, assuming the same kinematical (super-parallel) setup
as has been employed for the ($e,e'pp$) reaction displayed in Figure \ref{fig4}.
The resulting structure function for ($e,e'pn$) displayed in Figure \ref{fig6}
is almost an order of magnitude larger than for the corresponding ($e,e'pp$)
case. In ($e,e'pn$) reactions one has also to include the effects of MEC. Note,
however, that for the case considered the MEC contribution are smaller than the
correlation effects. This is due to a strong cancellation between the pion
seagull and the pion in flight contributions to the MEC. The dominating
contribution to the longitudinal response is again the correlation part.
Comparison with Figure \ref{fig4} demonstrates that the $pn$ correlations are
significantly larger than those for the $pp$ pairs. This supports our
conclusion from discussing the results of table \ref{tab1} that the pionic or
tensor correlations which are different for isospin $T=0$ and $T=1$ pairs play
an important role and are even more important than the central correlations,
which are independent of the isospin. 

More detailed results including the transversal structure function and the
effect of isobar currents have partly been published already in \cite{knoed}.
The discussion of further results is in preparation\cite{knoed1}

\section{Correlations in finite nuclei}

Various quite different approaches have been developed to determine correlations
in the nuclear wave function, which are beyond the mean-field or Hartree-Fock
approach. In the preceeding section we have employed a calculation of
correlation effects in terms of the Brueckner G-matrix. In this section we will
consider the so-called coupled cluster or ``$exp(S)$'' method. 
The basic features of the coupled cluster method have been described already in
the review article by K\"ummel et al.~\cite{kuem}. More recent developments and
applications can be found in~\cite{bish}. Here we will only present some basic
equations. The
many-body wave function of the coupled cluster or $\exp(S)$ method can be
written
\begin{equation}
\vert \Psi > =  \exp \left(\sum_{n=1}^A \hat S_n\right) \vert \Phi >\, .
\label{eq:exps}
\end{equation}
The state $\vert \Phi >$ refers to the uncorrelated model state, which we have
chosen to be a Slater determinant of harmonic oscillator functions with an
oscillator length $b$=1.72 fm, which is appropriate for the description of our
target nucleus $^{16}$O. The linked $n$-particle $n$-hole excitation operators
can be written
\begin{equation}
\hat S_n = \frac{1}{n!^2} \sum_{\nu_i\rho_i} <\rho_1\dots\rho_n\vert S_n \vert
\nu_1\dots \nu_n> a_{\rho_1}^\dagger \dots  a_{\rho_n}^\dagger a_{\nu_n} \dots
a_{\nu_1}\,.\nonumber
\end{equation}
Here and in the following the sum is restricted to oscillator states $\rho_i$
which are unoccupied in
the model state $\vert \Phi >$, while states $\nu_i$ refer to states which are
occupied in $\vert \Phi >$. For the application discussed here we assume the
so-called $S_2$ approximation, i.e.~we restrict the correlation operator in
(\ref{eq:exps}) to the terms with $\hat S_1$ and $\hat S_2$. 
One may introduce
one- and two-body wave functions
\begin{eqnarray}
\psi_1 \vert \nu_1> & = & \vert \nu_1 > + \hat S_1 \vert \nu_1 > \nonumber \\
\psi_2 \vert \nu_1 \nu_2 > & = & {\cal A} \, \psi_1 \vert \nu_1> \psi_1 \vert
\nu_2> + \hat S_2 \vert \nu_1 \nu_2 > \label{eq:psin}
\end{eqnarray}
with ${\cal A}$ denoting the operator antisymmetrizing the product of one-body
wave functions.
Using these definitions one can write the coupled equations for the evaluation
of the correlation operators $\hat S_1$ and $\hat S_2$ in the form
\begin{equation}
<\alpha \vert \hat T_1 \psi_1 \vert \nu > + \sum_{\nu_1} <\alpha \nu_1 \vert
\hat T_2 \hat S_2 + \hat V_{12} \vert \nu \nu_1 > = \sum_{\nu_1}
\epsilon_{\nu_1\nu} <\alpha \vert \psi_1 \vert \nu_1 >\,, \label{eq:hf}
\end{equation}
where $\hat T_i$ stands for the operator of the kinetic energy of particle $i$
and $\hat V_{12}$ is the two-body potential. Furthermore we introduce the
single-particle energy matrix defined by
\begin{equation}
\epsilon_{\nu_1\nu} = <\nu_1 \vert \hat T_1 \vert \nu > + \sum_{\nu'} < \nu_1
\nu' \vert \hat V_{12} \psi_2 \vert \nu \nu'> \nonumber
\end{equation}
The Hartree-Fock type equation (\ref{eq:hf}) is coupled to a two-particle
equation of the form
\begin{eqnarray}
0 & = & <\alpha \beta \vert \hat Q \left[ (\hat T_1 + \hat T_2)\hat S_2 +
\hat
V_{12}\psi_2 + \hat S_2 \hat P \hat V_{12} \psi_2 \right] \vert \nu_1 \nu_2 >
\nonumber \\
&& \qquad
- \sum_{\nu} \left( <\alpha \beta \vert \hat S_2 \vert \nu \nu_2 >
\epsilon_{\nu\nu_1} + <\alpha \beta \vert \hat S_2 \vert \nu_1 \nu >
\epsilon_{\nu\nu_2} \right)  \label{eq:s2}
\end{eqnarray}
In this equation we have introduced the Pauli operator $\hat Q$ projecting on
two-particle states, which are not occupied in the uncorrelated model state
$\vert \Phi >$ and the projection operator $\hat P$, which projects on
two-particle states, which are occupied. If for a moment we ignore the term in
(\ref{eq:hf}) which is represented by the operators $\hat T_2 \hat S_2$ and
also the term in (\ref{eq:s2}) characterized by the operator  $\hat S_2 \hat P
\hat V_{12}$ the solution of these coupled equations corresponds to the
Brueckner-Hartree-Fock approximation and we can identify the matrix elements of
$\hat V_{12} \psi_2$ with the Brueckner $G$-matrix. Indeed the effects of these
two terms are rather small and we have chosen the coupled cluster approach
mainly because it provides directly correlated two-body wave functions (see
eq.\ref{eq:psin}). More details about the techniques which are used to solve the
coupled cluster equations can be found in \cite{zab1,carlpn}.

As an example we would like to present the effects of correlations on the
two-body density obtained by removing two protons from oscillator $p_{1/2}$
states, coupled to total angular momentum $J=0$ and isospin $T=1$
\begin{equation}
\left| <\vec r_1 \vec r_2 \vert \psi_2 \vert p_{1/2},p_{1/2}\,J=0,T=1 >\right|^2
\label{eq:2pdens}
\end{equation}

\begin{figure}[tb]
\begin{center}
\epsfig{file=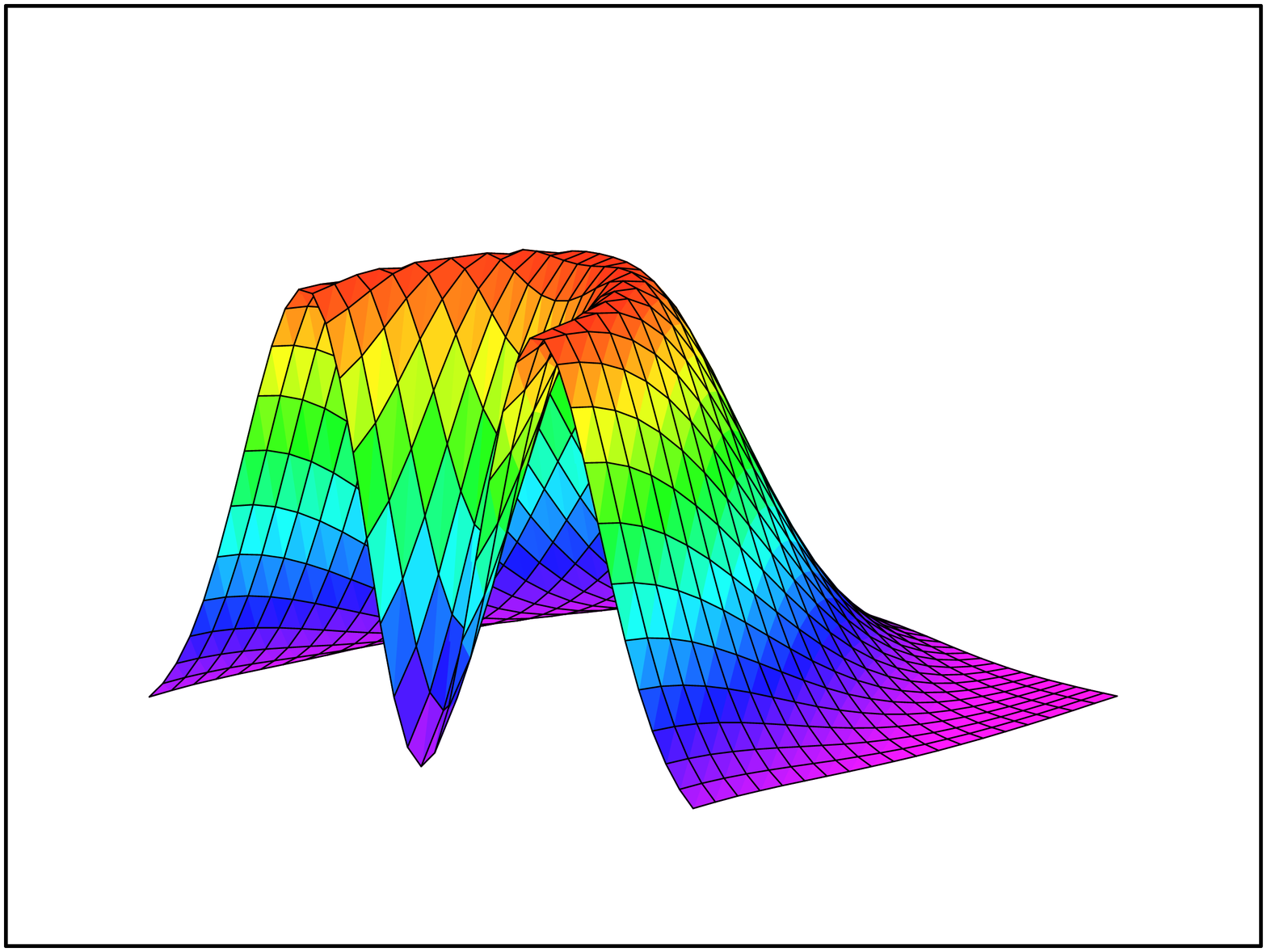,scale=0.25,angle=270}

\epsfig{file=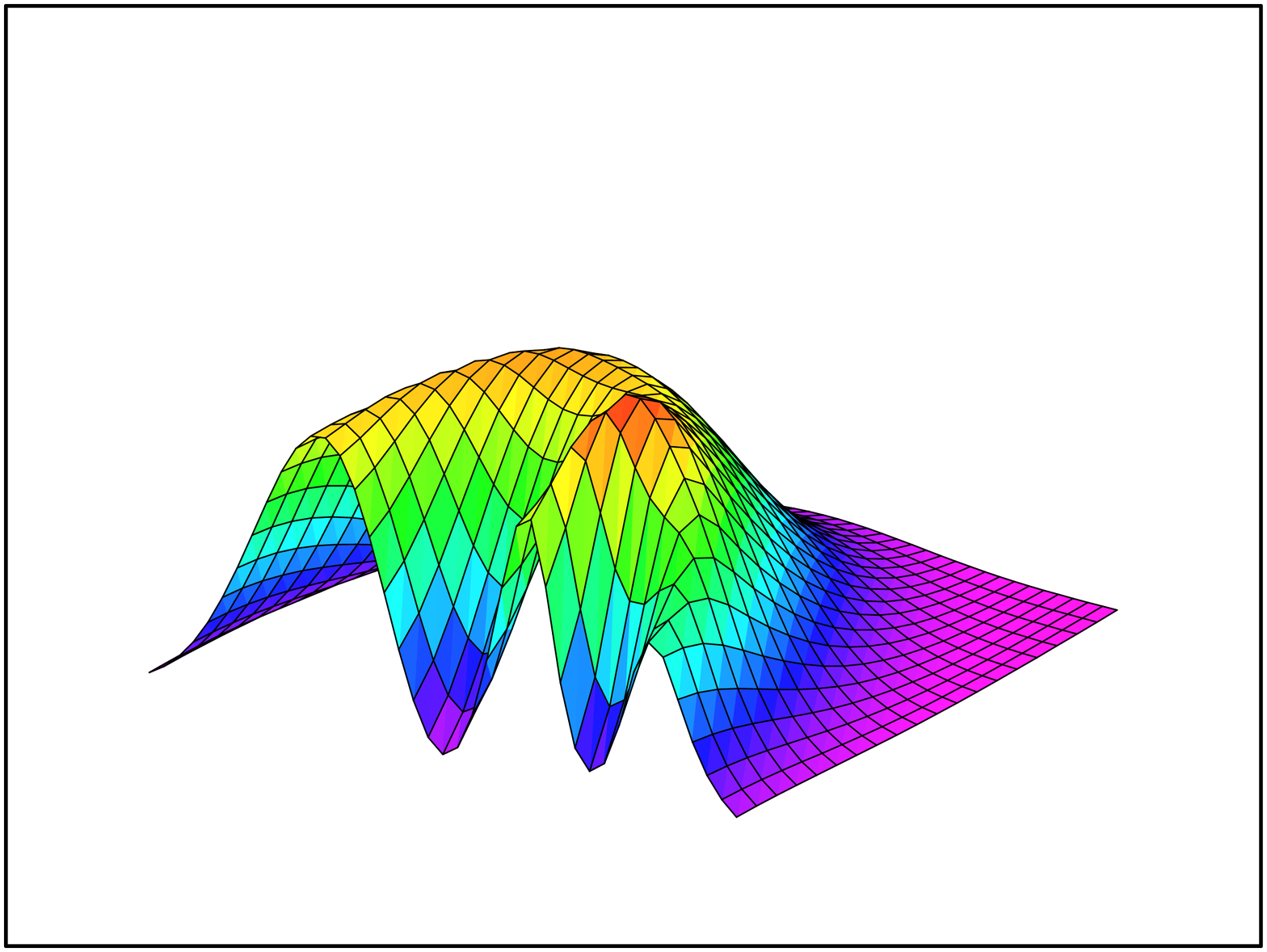,scale=0.25,angle=270}
\end{center}
\caption{\label{fig7}
Two-body density according to (\protect\ref{eq:2pdens}) for a fixed vector $\vec
r_1$ as a function of $\vec r_2$. The upper part displays the result without
correlations ($\hat S_2 = 0$), while correlations are included in the lower part
of the figure. Further description in the text}
\end{figure}

In Figure \ref{fig7} this two-body density is displayed for
a fixed $\vec r_1 = (x_1=0, y_1=0, z_1=2\ {\rm fm})$ as a function of $\vec
r_2$, restricting the presentation to the $x_2,z_2$ half-plane with ($x_2 >0, y_2
=0$). The upper part of this figure displays the two-body density without
correlations ($\hat S_2 = 0$). One observes that the two-body density, displayed
as a function of the position of the second particle $\vec r_2$ is not affected
by the position of the first one $\vec r_1$. Actually, the two-body density
displayed is equivalent to the one-body density. This just reflects the feature
of independent particle motion. If correlation effects are included, as it is
done in the lower part of Figure \ref{fig7}, one finds a drastic reduction of 
the two-body density at $\vec r_2 = \vec r_1$ accompanied by a slight 
enhancement at medium separation between $\vec r_1$ and $\vec r_2$. 

\begin{figure}[tb]
\begin{center}
\epsfig{file=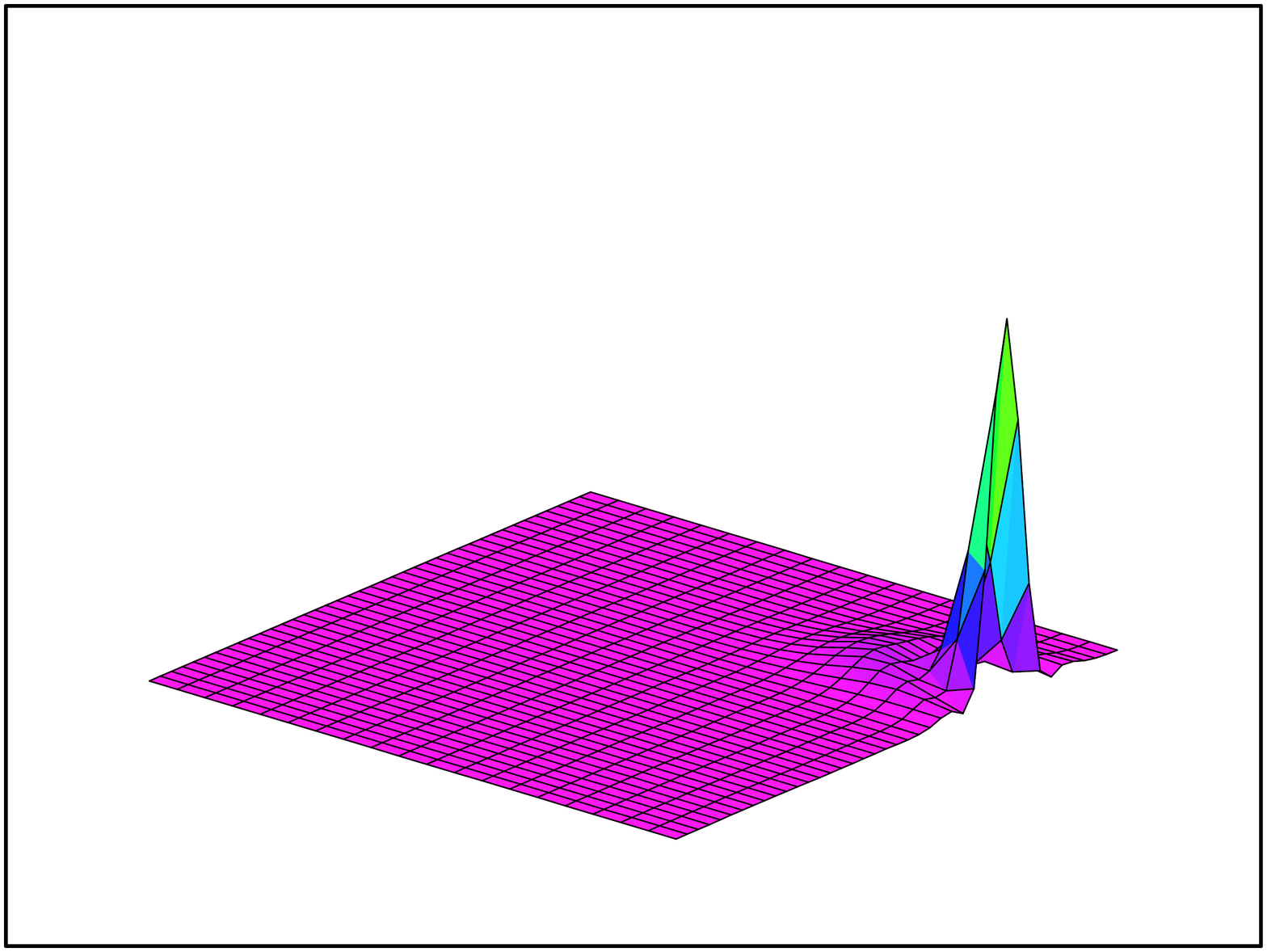,scale=0.25,angle=270}

\epsfig{file=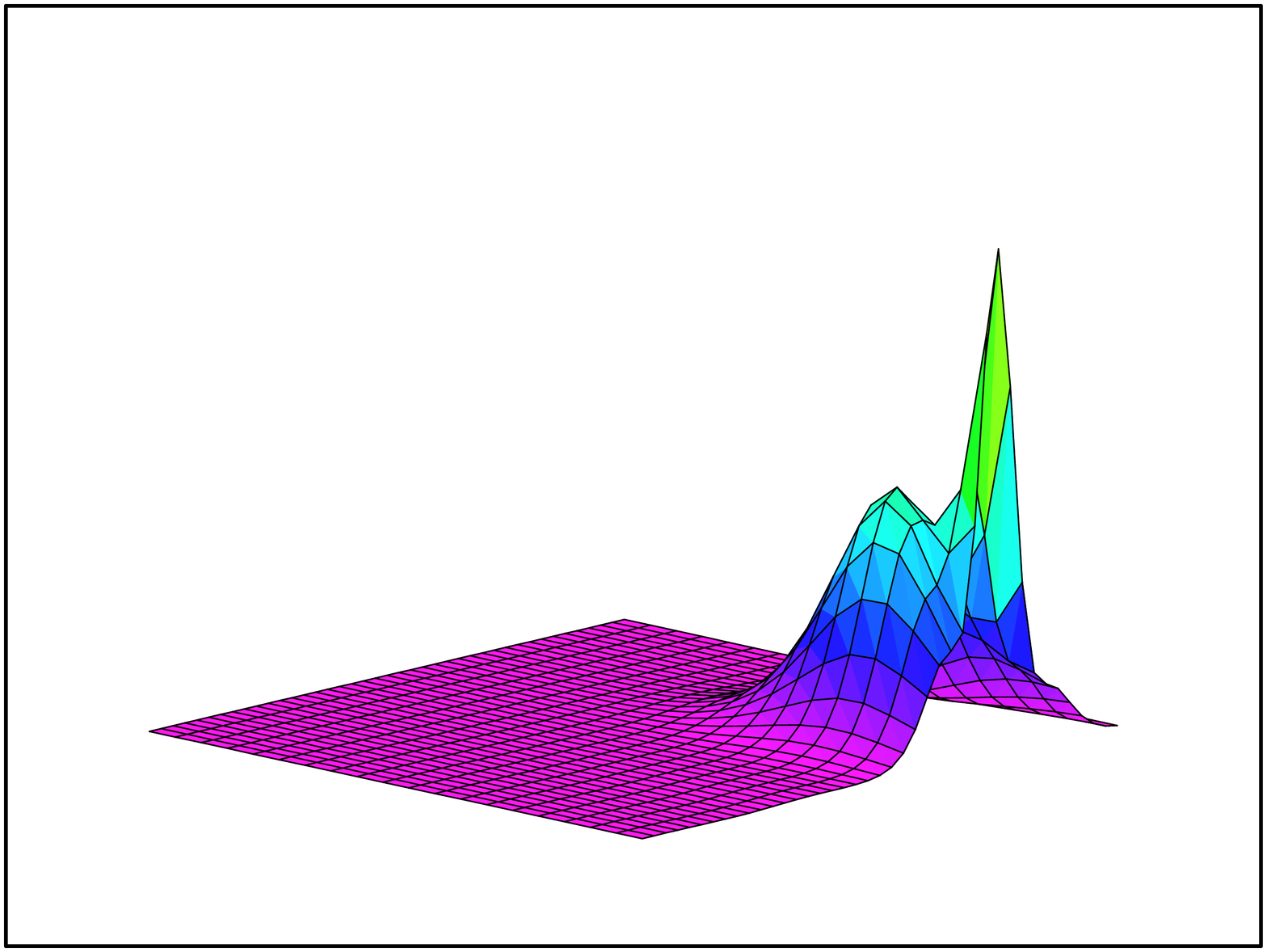,scale=0.25,angle=270}
\end{center}
\caption{\label{fig8}
Correlation density according to (\protect\ref{eq:2pdens}) with $\psi_2$
replaced by $\hat S_2$ for a fixed vector $\vec
r_1$ as a function of $\vec r_2$. The upper part displays the result for a
$J=0,T=1$ pair (pp) while the lower part refers to the removal of a $J=1,T=0$
(pn) pair.}
\end{figure}

In order to amplify the effect of correlations, Figure \ref{fig8} displays the
corresponding correlation densities (i.e.~replace $\psi_2$ by $\hat S_2$, see
also (\ref{eq:psin})). While the upper part shows the correlation density for
the removal of a proton-proton pair, the corresponding density for a
proton-neutron pair is displayed in the lower part. Comparing these figures one
sees that that the $pn$ correlations are significantly stronger than the $pp$
correlations. This is mainly due to the presence of pionic or tensor
correlations in the case of the $pn$ pair. Figure \ref{fig8} also exhibits quite
nicely the range of the correlations. This range is short compared to the size
of the nucleus even in the case of the $pn$ correlations. All results displayed
in this section have been obtained using the Argonne V14 potential for the NN
interaction\cite{v14}.
 
\section{Two nucleon knockout on $^{16}O$}

The coincidence cross section for the reaction induced by an electron
with momentum $\p_{0}$ and energy $E_{0}$, with $E_{0}=|\p_{0}|=p_{0}$, where
two nucleons, with momenta $\p'_{1}$, and $\p'_{2}$
and energies $E'_{1}$ and $E'_{2}$, are ejected from a nucleus is given, in the
one-photon exchange approximation and after
integrating over $E'_{2}$, by~\cite{GP}
\begin{equation}
\frac{{\mathrm d}^{8}\sigma}{{\mathrm d}E'_{0}{\mathrm d}\Omega
{\mathrm d}E'_{1}{\mathrm d}\Omega'_{1}
{\mathrm d}\Omega'_{2}} = K \Omega_{\mathrm f} f_{\mathrm{rec}}
|j_\mu J^\mu|^2 .
\label{eq:cs}
\end{equation}
In Eq.~(\ref{eq:cs}) $E'_{0}$ is the energy of the scattered electron with
momentum $\p'_{0}$, $K = e^4{p'_{0}}^2/4\pi^2 Q\,^4$ where
$Q^2 = \q\,^2 - \omega^2$, with $\omega = E_{0} - E'_{0}$ and
$\q = \p_0 - \p'_0$, is the four-momentum transfer. The quantity
$\Omega_{\mathrm f} = p'_{1} E'_{1} p'_{2} E'_{2}$ is the phase-space
factor and integration over $E'_{2}$ produces the recoil factor
\begin{equation}
f_{\mathrm{rec}}^{-1} = 1 - \frac{E'_{2}}{E_{\mathrm B}} \, \frac{\p'_{2}\cdot
\p_{\mathrm B}}{|\p'_{2}|^2},
\end{equation}
where $E_{\mathrm B}$ and $\p_{\mathrm B}$ are the energy and momentum of the
residual nucleus. The cross section is given by the square of the scalar
product of the relativistic electron current $j^\mu$ and of the nuclear
current $J^\mu$, which is given by the Fourier transform of the transition
matrix elements  of the charge-current density operator between initial and
final nuclear states
\begin{equation}
J^\mu (\q) = \int < \Phi_{\mathrm{f}} | \hat{J}^\mu(\r) |\Phi_{\mathrm{i}} >
{\mathrm{e}}^{\,{\mathrm{i}}{\footnotesize \q} \cdot
{\footnotesize \r}} {\mathrm d}\r.
\label{eq:jm}
\end{equation}

If the residual nucleus is left in a discrete eigenstate of its Hamiltonian,
i.e. for an exclusive process, and under the assumption of a direct knockout
mechanism, Eq.~(\ref{eq:jm}) can be written as~\cite{GP}
\begin{eqnarray}
J^\mu(\q) & = &  \int
\phi_{\mathrm{f}}^{*}(\r_{1}\ss_{1},\r_{2}\ss_{2})
J^\mu(\r,\r_{1}\ss_{1},\r_{2}\ss_{2})\phi_{\mathrm{i}}
(\r_{1}\ss_{1},\r_{2}\ss_{2}) \nonumber \\
& & \times \,{\mathrm{e}}^{\,{\mathrm{i}}{\footnotesize \q} \cdot
{\footnotesize \r}} {\mathrm d}\r{\mathrm d}\r_{1} {\mathrm d}\r_{2}
{\mathrm d}\ss_{1} {\mathrm d}\ss_{2} . \label{eq:jq}
\end{eqnarray}

Eq.~(\ref{eq:jq}) contains three main ingredients: the two-nucleon overlap
integral $\phi_{\mathrm{i}}$, the nuclear current $J^\mu$ and the final-state
wave function $\phi_{\mathrm{f}}$.

In the model calculations the final-state wave function $\phi_{\mathrm {f}}$
includes the interaction of each one of the two outgoing nucleons with the
residual nucleus while their mutual interaction, which we have discussed as FSI
in the preceeding section is here neglected. Therefore, the
scattering state is written as the product of two uncoupled single-particle
distorted wave functions, eigenfunctions of a complex phenomenological optical
potential which contains a central, a Coulomb and a spin-orbit term.

The nuclear current operator in Eq.~(\ref{eq:jq}) is the sum of a one-body
and a two-body part. In the one-body part convective and spin currents are
included. As discussed already in section 3, the two-body current includes, 
the seagull and pion-in-flight diagrams and the diagrams
with intermediate isobar configurations.

\begin{figure}[tb]
\begin{center}
\epsfig{file=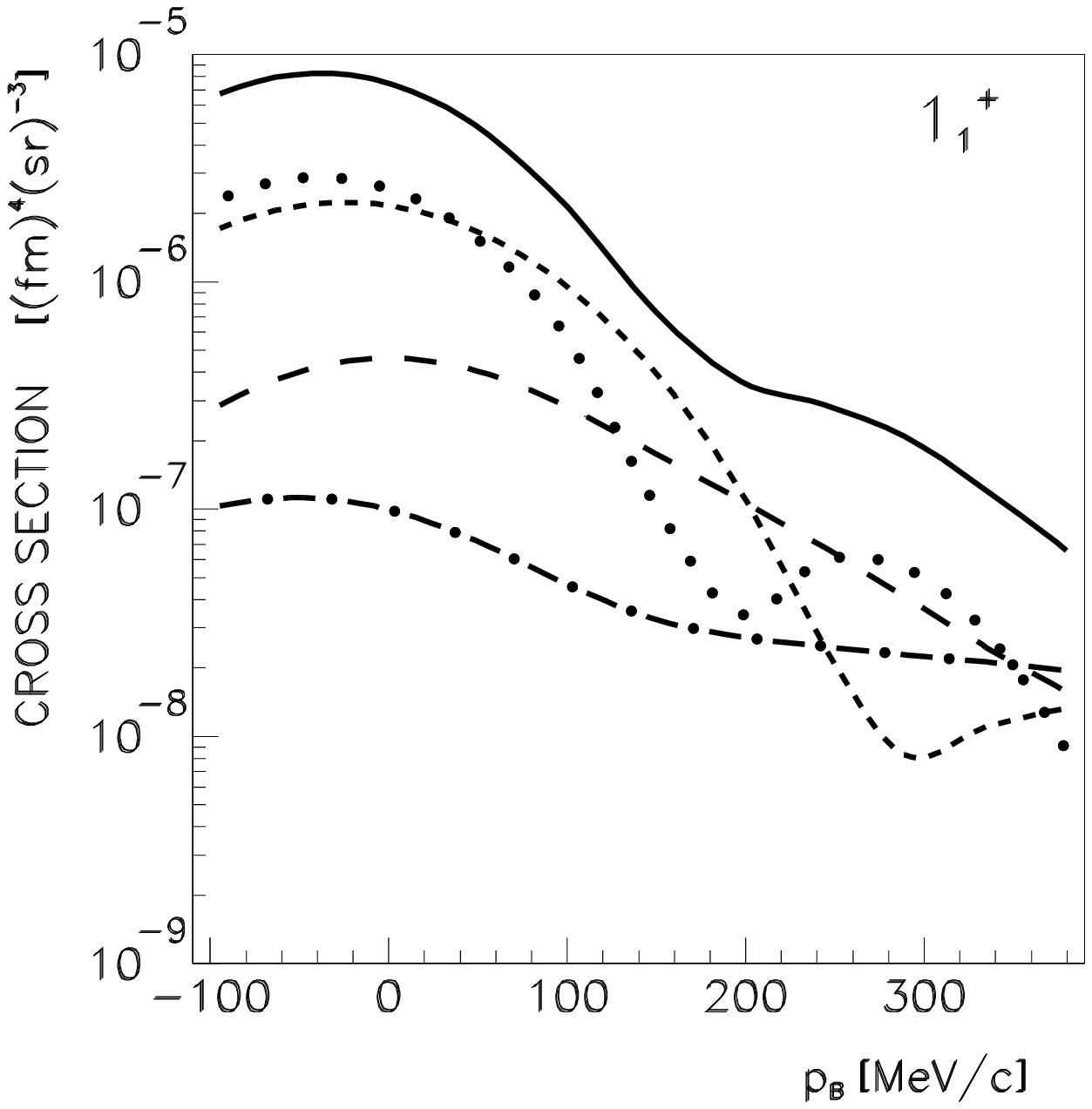,scale=0.50}
\end{center}
\caption{\label{fig9}
The differential cross section  of the $^{16}$O(e,e$'$pn)
reaction as a function of the recoil momentum $p_{\mathrm{B}}$ for the
transition to the $1^+_1$ ground state of $^{14}$N ($E_{2\mathrm{m}} = 22.96$
MeV), in the super-parallel kinematics with $E_{0} = 855$ MeV, and
$\omega = 215$ MeV $q = 316$ MeV/$c$. The recoil-momentum distribution is
obtained changing the kinetic energies of the outgoing nucleons. Separate
contributions of the one-body, seagull, pion-in-flight and $\Delta$-current are
shown by the dotted, short-dashed, dot-dashed and long-dashed lines,
respectively.  Positive (negative) values of
$p_{\mathrm{B}}$ refer to situations where
${\mbox{\boldmath $p$}}_{\mathrm{B}}$ is parallel (antiparallel) to
${\mbox{\boldmath $q$}}$.
}
\end{figure}

The two-nucleon overlap integral $\phi_{\mathrm{i}}$ contains the information
on nuclear structure and allows one to write the cross section in terms of the
two-hole spectral function.
For a discrete final state of the $^{14}$N nucleus, with angular momentum
quantum number $J$, the state $\phi_{\mathrm{i}}$ is expanded in terms of the
correlated two-hole wave functions defined in the preceeding section as
\begin{eqnarray}
\phi_{\mathrm{i}}^{JT}(\vec r_{1}{\ss}_{1},
{\vec r}_{2}{\ss}_{2}) & = & \sum_{\nu_1 \nu_2}
 a^{JT}_{\nu_1\nu_2} <\vec r_{12},\vec R ,{\ss}_1,{\ss}_2\vert \psi_2 \vert 
\nu_1 \nu_2 JT> \label{eq:ppover}
\end{eqnarray}
The expansion coefficients $a^{JT}_{\nu_1\nu_2}$ are determined from a
configuration mixing calculations of the two-hole states in $^{16}$O, which can
be coupled to the angular momentum and parity of the requested state. The
residual interaction for this shell-model calculation is also derived from the
Argonne V14 potential and corresponds to the Brueckner G-matrix. Note that these
expansion coefficients $a^{JT}_{\nu_1\nu_2}$ account for the global or 
long-range
structure of the specific nuclear states, while the information on short-range
correlations is already contained in $<\vec r_{12},\vec R ,{\ss}_1,{\ss}_2\vert
 \psi_2 \vert \nu_1 \nu_2 JT>$.

\begin{figure}[tb]
\begin{center}
\epsfig{file=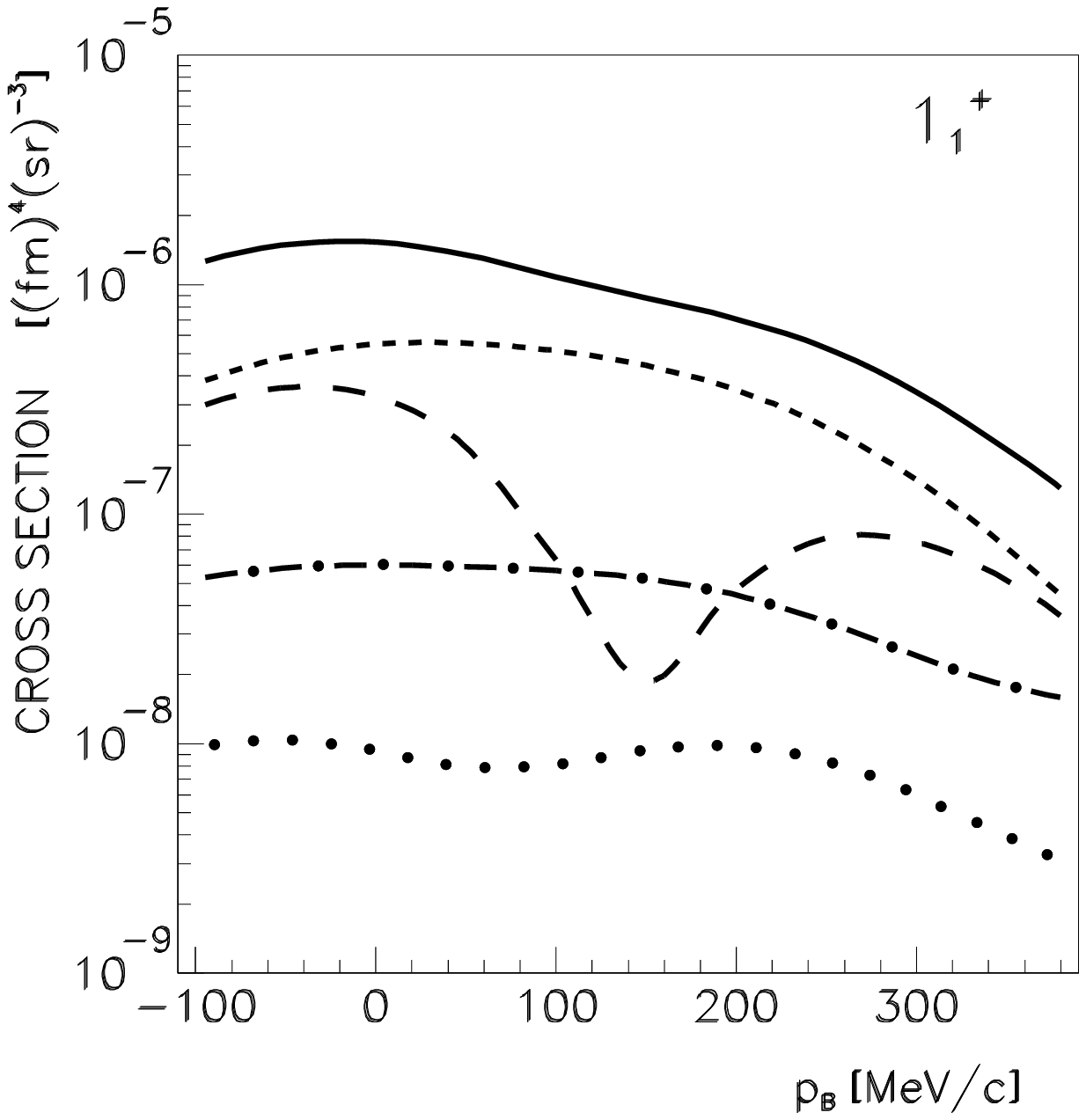,scale=0.50}
\end{center}
\caption{\label{fig10}
The same as Fig.~\ref{fig9} but with a simpler approach 
for the two-nucleon overlap.}
\end{figure}

Results for the cross section of exclusive ($e,e'pn$) reactions on $^{16}$O
leading to the ground state of $^{14}N$ are displayed in Figure \ref{fig9}.
The calculations have been performed in the super-parallel kinematic, which we
already introduced before. The kinematical parameters correspond to those
adopted in a recent $^{16}$O(e,e$'$pp)$^{14}$C experiment at
MAMI~\cite{Rosner}. In order to allow a direct comparison of ($e,e'pp$) with
($e,e'pn$) experiments, the same setup has been proposed for 
the first experimental study of
the $^{16}$O(e,e$'$pn)$^{14}$N reaction~\cite{MAMI}. This means that we assume 
an energy of the incoming electron $E_0 =855$ MeV, electron 
scattering angle $\theta=
18^{\mathrm{o}}$, $\omega = 215$ MeV and $q= 316$ MeV/$c$. The proton is
emitted parallel and the neutron antiparallel to the momentum transfer $\q$.

Separate contributions of the different terms of
the nuclear current are shown in the figure and compared with the total cross
section\cite{carlpn}. The contribution of the one-body current, entirely due to
correlations, is large. It is of the same size as that of the pion seagull
current. The contribution of the $\Delta$-current is much smaller at lower
values of $p_{\mathrm{B}}$, whereas for values of $p_{\mathrm{B}}$ larger than
100 MeV/$c$ it becomes comparable with that of the other components.
It is worth noting the the total cross section is about an order of magnitude
larger than the one evaluated for the corresponding ($e,e'pp$)
experiment\cite{gius1}. This confirms our finding about the relative cross
sections for $pp$ and $pn$ knock out, which we have observed already in section
3.

\begin{figure}[tb]
\begin{center}
\epsfig{file=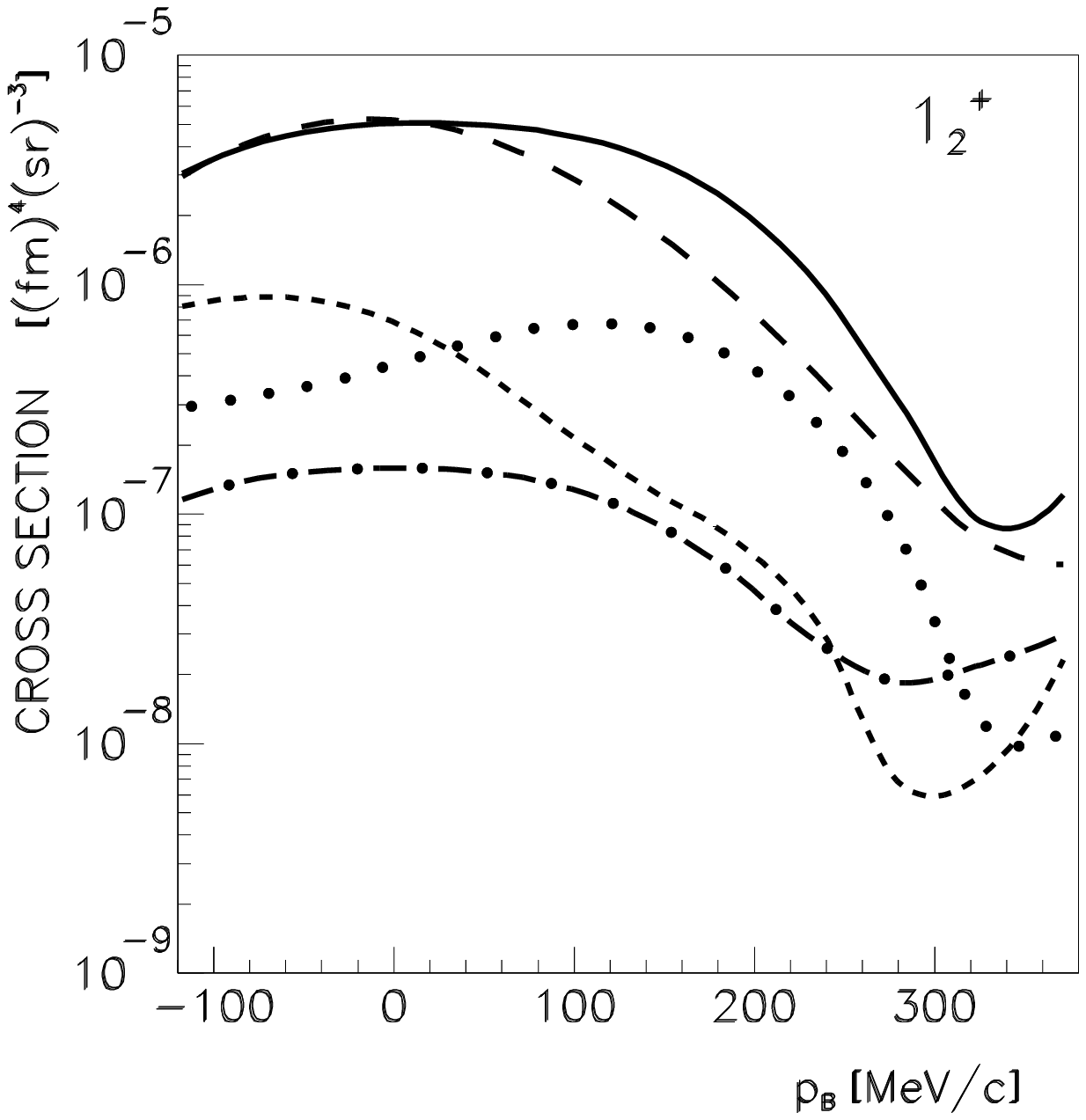,scale=0.50}
\end{center}
\caption{\label{fig11}
The same as Fig.~\ref{fig9} for the transition to the $1^+_2$
state of $^{14}$N ($E_{2\mathrm{m}} = 26.91$ MeV).}
\end{figure}

In Fig.~\ref{fig10} the same quantities as in Fig.~\ref{fig9} are shown,
but the two-nucleon overlap has been calculated with the simpler prescription
of correlations, i.e. by the product of the pair function of the shell
model, described for $1^+_1$ as a pure ($p_{1/2}$)$^{-2}$ hole, and of a
Jastrow type central and state independent correlation function.
The large differences between the cross sections in Figs.~\ref{fig9}
and~\ref{fig10} indicate that a refined description of the two-nucleon
overlap, involving a careful treatment of both aspects related to nuclear
structure and NN correlations, is needed to give reliable predictions of the
size and the shape of the ($e,e'pn$) cross section.

The cross sections for the transition to the excited 1$^+_2$ state are 
displayed in
Figs.~\ref{fig11}. The two-nucleon overlap function for
this state contains the same components in terms of relative and c.m. wave
functions and the same defect functions as for the 1$^+_1$ ground state, but
they are weighed with different amplitudes $a^{J}_{\nu_1\nu_2}$ in
Eq.~(\ref{eq:ppover}). In practice the two overlap functions have different
amplitudes for $p_{1/2}$ and $p_{3/2}$ holes. This has the
consequence that the cross
sections in Figs.~\ref{fig9} and~\ref{fig11} have a different shape and
are differently affected by the various terms of the nuclear current.
So transition to various states probe the ingredients of the transient
matrix elements in different ways. More details will be presented in the
contribution of Carlotta Giusti\cite{carlcon}.

\section{Conclusion}

It has been the aim of this contribution to demonstrate that exclusive
($e,e'NN$) reactions are sensitive to NN correlations and therefore sensitive to
the NN interaction in the nuclear medium at short inter-nucleon distances. The
careful study and analysis of these reactions is a challenge for experimental
but also theoretical efforts. In particular it should be pointed out:

\begin{itemize}
\item $pp$ as well as $pn$ knock-out experiments should be performed. The cross
sections for $pn$ knock-out are significantly larger than for corresponding $pp$
emission. This is partly due to the meson-exchange-current (MEC) contributions
for the charged mesons, which is absent in $pp$ knock-out. The difference,
however, also reflects the isospin dependence of nuclear correlations. While
the study of ($e,e'pp$) mainly explores the short-range central correlations,
the corresponding ($e,e'pn$) experiments also probe tensor correlations.
\item Effects of Final State Interaction (FSI) are non-negligible. Most of the
studies up to now consider FSI effects only in a mean field approach. It must be
emphasized, however, that the residual interaction between the two ejected
nucleons has a non-negligible effect as well. This is even true, when the two
nucleons are emitted `back - to - back'. FSI effects, however, get much more
important for other final states.
\item All contributions to the ($e,e'NN$) cross section should be determined in
a consistent way. In order to separate the various contributions, one should try
to separate the various structure functions (longitudinal and transverse). One
may also take advantage of the fact that transitions to various final states in
the residual nucleus probe the different contributions differently.
\item The super-parallel kinematic seems to be quite appropriate for the study of
correlation effects.  
\end{itemize}

\noindent{\bf Acknowledgments}\\[1ex]
The results, which have been presented here, have been obtained in collaborations
with many colleagues. In particular I would like to mention the PhD students 
Daniel Kn\"odler, Markus Stauf and Stefan Ulrych. Furthermore, I would like to
thank K.~Allaart, K.~Amir-Azimi-Nili, P.~Czerski, W.H.~Dickhoff, C.~Giusti,
F.D.~Pacati, A.~Polls and J.~Udias. This work has been supported by grants from 
the DFG (SFB 382,  GRK 135 and  Wa728/3).

\end{document}